# Quasiparticle interference and charge order in a heavily overdoped non-superconducting cuprate

Xintong Li[1], Ying Ding[2], Chaocheng He[3], Wei Ruan[1], Peng Cai[1], Cun Ye[1], Zhenqi Hao[1], Lin Zhao[2], Xingjiang Zhou[2,4], Qianghua Wang[3,5], Yayu Wang[1,4†]

[1]*State Key Laboratory of Low Dimensional Quantum Physics, Department of Physics, Tsinghua University, Beijing 100084, China*

[2]*Beijing National Laboratory for Condensed Matter Physics, Institute of Physics, Chinese Academy of Sciences, Beijing 100190, P. R. China*

[3]*National Laboratory of Solid State Microstructures, Nanjing University, Nanjing, 210093, P. R. China*

[4]*Collaborative Innovation Center of Quantum Matter, Beijing, China*

[5]*Collaborative Innovation Center of Advanced Microstructures, Nanjing University, Nanjing 210093, China*

[†] Email: yayuwang@tsinghua.edu.cn

**One of the key issues in unraveling the mystery of high $T_C$ superconductivity in the cuprates is to understand the normal state outside the superconducting dome. Here we perform scanning tunneling microscopy and spectroscopy measurements on a heavily overdoped, non-superconducting $(Bi,Pb)_2Sr_2CuO_{6+\delta}$ cuprate. Spectroscopic imaging reveals dispersive quasiparticle interferences and the Fourier transforms uncover the evolution of momentum space topology. More interestingly, we observe nanoscale patches of static charge order with $\sqrt{2} \times \sqrt{2}$ periodicity. Both the dispersive quasiparticle interference and static charge order can be qualitatively explained by theoretical calculations, which reveal the unique electronic structure of strongly overdoped cuprate.**

The superconducting (SC) state of high $T_C$ cuprates exists within a "dome" in the phase diagram and disappears both in the severely underdoped and heavily overdoped limits. Because the cuprates are widely believed to be doped Mott insulators [1], the underdoped regime near the parent compound has been extensively studied by various experimental techniques, which have revealed highly unusual phenomena such as the pseudogap phase [2] and complex charge/spin orders [3-14]. On the contrary, the heavily overdoped regime is much less explored because it is generally considered to be a rather conventional Fermi liquid (FL) state. This point has been illustrated by the crossover from a non-FL-like linear temperature ($T$) dependent resistivity at optimal doping to the $T^2$ dependent resistivity characteristic of Landau FL in the heavily overdoped regime [15-19], as well as quantum oscillation experiments revealing a single hole-like Fermi surface (FS) [20, 21]. Because the physics of the FL is well-understood, the heavily overdoped limit can actually serve as another valid starting point, presumably more accessible than the Mott insulator limit, for understanding the origin of superconductivity in cuprates.

Previous experiments on overdoped cuprates have revealed a number of important features regarding the electronic structure. Angle-resolved photoemission spectroscopy (ARPES) shows a FS topology transition from a $(\pi,\pi)$-centered hole-like pocket to a $(0,0)$-centered electron-like pocket [22-24]. In single band tight binding model [25, 26], the change of FS topology in two-dimension should be accompanied by a logarithmic divergence of electron density of states (DOS) known as the Van Hove singularity (VHS) [27]. Recent scanning tunneling microscopy (STM) experiments provide direct evidence for VHS in heavily overdoped cuprate, as well as the existence of pseudogap [26]. However, it is still unclear what the main difference is, from the electronic structure and electronic order point of view, between the FL and SC states across the phase boundary in the overdoped side. Especially, the charge order phenomenon, which is ubiquitous in underdoped cuprates and entangles intricately with superconductivity [7, 28-30], has been mostly neglected in the heavily overdoped non-SC regime of the phase diagram.

In order to elucidate the electronic structure and electronic order in the overdoped regime outside the SC dome, here we perform STM studies on a heavily overdoped, non-SC Bi$_2$-

$_x$Pb$_x$Sr$_2$CuO$_{6+\delta}$ (Pb-Bi2201) cuprate. Tunneling spectroscopy reveals the VHS feature and its evolution into the pseudogap phase, and the dispersive quasiparticle interference (QPI) patterns reveal the change of FS topology. More remarkably, we observe nanoscale patches of static charge orders with $\sqrt{2} \times \sqrt{2}$ periodicity. The possible origin of the charge order and its implications to the superconductivity will be discussed.

## Results

**Spatially resolved tunneling spectroscopy.**

The Pb doped Bi2201 is chosen because it can be overdoped into the non-SC regime and has an ideal cleaved surface. High-quality Pb-Bi2201 single crystals are grown by the traveling solvent floating zone method and the $T_C$ of the as-grown sample is about 3 K [31]. The non-SC sample studied in this work is obtained by annealing the as-grown sample in high pressure O$_2$ (~ 80 atm) at 500 ℃ for 7 days to further increase the hole density. It exhibits no sign of superconductivity down to 2 K. Figure 1(a) depicts the schematic electronic phase diagram, and the red arrow shows the approximate location of the non-SC sample. The Pb-Bi2201 crystal is cleaved in the ultrahigh vacuum chamber at $T$ = 77 K, and is then transferred into the STM chamber with the sample stage cooled to $T$ = 5 K. STM topography is taken in the constant current mode with an electro-chemically etched tungsten tip, which has been treated and calibrated on a clean Au(111) surface [32]. The $dI/dV$ (differential conductance) spectra are collected by using a standard lock-in technique with modulation frequency $f$ = 423 Hz. All the data reported here are taken at $T$ = 5 K.

Shown in Fig. 1(b) is the exposed (Bi,Pb)O surface topography of a non-SC sample, which shows a regular square lattice. The structural supermodulation usually observed in Bi-based cuprates is suppressed by Pb doping [28, 33]. Around 13% of the atomic sites are bright spots, which is consistent with the 12.2% Pb substitution of Bi determined by the sample growth condition [28, 33]. There are spatial inhomogeneities with typical size around a few nanometers, which presumably result from the non-uniform distribution of local hole density [34, 35].

The local electronic structure is probed by *dI/dV* spectroscopy, which is approximately proportional to the electron DOS. Figure 1(c) displays a series of representative spectra taken at various locations indicated by the corresponding colored dots in Figure 1(b). The spectra exhibit significant but yet systematic variations. Roughly speaking there are two types of spectra, one with a prominent peak near the Fermi energy ($E_F$) and the other with a DOS suppression around $E_F$ that is reminiscent of the pseudogap. In Fig. 1(d) we show that the peaks in *dI/dV* can be fitted well by a simple function $a + b \log|E-E_{VHS}|$ with $E_{VHS}$ denoting the peak position, which is consistent with the spectrum due to the presence of VHS [36]. The spectra with DOS suppression are quite similar to that in OD cuprate with lower hole density in the overdoped SC regime [37, 38]. The spatially averaged *dI/dV* spectrum in Fig. 1(e) exhibits a DOS peak around $E_F$, revealing that statistically the dominant spectral feature in this sample is the VHS-type. The spatial variations of the spectra reflect that the VHS-type spectra gradually evolve into the pseudogap-type with reduced doping, which is consistent with the expected band structure evolution of overdoped cuprates.

**The *dI/dV* maps and dispersive QPI patterns.**

Next we will focus on the electronic structure and electronic order in this sample. Figures 2(a)-(c) display three representative *dI/dV* maps measured at three different sample biases, $V_b$ = 40 mV, -10 mV and -60 mV on the same field of view as Fig. 1(b). The different spatial patterns at different energies indicate the existence of strong dispersions due to the quantum interference of coherent quasiparticles, or QPI [39, 40]. Figures 2(d) to 2(f) display the Fourier transform (FT) maps of the three *dI/dV* maps, in which the four sharp Bragg peaks with lattice wavevector $q_0 = 2\pi/a_0$ ($a_0$ is the lattice constant) can be used to extract the QPI wavevector *q*. With varied energies, the main changes of the FT-QPI patterns occur along the antinodal direction pointing to the lattice Bragg peaks. The two antinodal QPI branches gradually merge together with decreasing bias voltage (Fig. 2(e)), develop a closed pocket centered at (0, 0), and then shrink further with lowering energy. The overall energy dependence of the QPI patterns are similar to that in an overdoped Bi2201 with $T_C$ = 15 K reported previously [38].

The calculated QPI patterns (Figs. 2(g)-(i)) and constant energy contours (CECs) in the momentum space (Figs. 2(j)-(l)) are presented. The QPI patterns are calculated from a minimal tight-binding model following previous works [39-44]. The generalized T-matrix formalism is used to get the local density of states (LDOS) in the presence of impurities, and the QPI pattern is calculated as the power spectrum of the LDOS. The representative CEC and corresponding QPI pattern is calculated by tuning the energy (or bias voltage in the STM experiments). With decreasing bias voltage, the CEC evolves from hole pocket around zone corner to VHS at a saddle point at ($\pi$, 0) resembling Fig. 1(d), and finally to an electron pocket around the zone center. The QPI patterns calculated by the octat model based on these CECs agree well with the experimental ones at these energies. The energy dependent CEC topology transition has the same trend as the hole-concentration dependent FS topology transition detected by ARPES in overdoped Pb-Bi2201 [23].

**The static $\sqrt{2} \times \sqrt{2}$ charge order structure.**

In addition to the dispersive QPI features, a more important, and totally unexpected feature revealed by the *dI/dV* maps are the existence of non-dispersive structure when we examine the low energy *dI/dV* maps. As displayed by the dashed squares in Fig. 3(a), the DOS map at zero bias exhibits nanoscale patches of short-range charge orders with a 45-degree rotation with respect to the atomic lattice. This feature has never been observed in cuprates before [7, 26, 29, 38]. The *dI/dV* maps obtained at different bias energies indicate that the charge order is more pronounced around $E_F$, and is visible over the entire energy range. To inspect its fine structures, we show in Figs. 4(b) and 4(c) the zoomed-in topographic and *dI/dV* maps acquired at $V_b = 0$ mV on the area enclosed by the green dashed square in Fig. 3(a). It is clearly illustrated from the comparison of these two maps that the charge order locally imposes a commensurate $\sqrt{2} \times \sqrt{2}$ superstructure on the original atomic lattice. Moreover, the charge order patterns of different patches in Fig. 3(a) do not align with each other. The lack of long-range order indicates that the charge order may be affected by local impurities or inhomogeneous distribution of hole concentration.

We gain more insight into the charge order by investigating its dependence on the bias voltage. Depicted in Figs. 4(d)-4(g) are the *dI/dV* maps of a small charge ordered patch (marked by the red dashed square in Figs. 4(b) and 4(c)) acquired at $V_b$ = 0 mV, -5 mV, -10 mV and -20 mV, demonstrating that this charge order keeps a commensurate periodicity without any dispersion within the energy range from -20 mV to 0 mV. This suggests that the $\sqrt{2} \times \sqrt{2}$ pattern is a static charge order, in sharp contrast to the dispersive QPI patterns.

## Discussion

Previous STM studies in underdoped cuprates have revealed the ubiquitous existence of charge order with wavevector around 4 $a_0$ along the Cu-Cu bond direction [7, 30]. However, the $\sqrt{2} \times \sqrt{2}$ charge order reported here has never been observed before in cuprates. In fact, the issue of charge order in heavily overdoped cuprates has been mostly neglected so far, and previous study on an overdoped Bi2201 with $T_C$ = 15 K did not observe such charge order [38]. Therefore, the $\sqrt{2} \times \sqrt{2}$ charge order could be a unique feature of the non-SC regime of the cuprate phase diagram, and may reveal key information regarding how superconductivity is suppressed by strong overdoping. The main questions regarding the observed charge order are its origin and implications to the SC phase. Below we present a possible mechanism to account for the $\sqrt{2} \times \sqrt{2}$ charge order in the strongly overdoped regime.

A likely cause for the charge order is the competition between onsite Coulomb repulsion $U$ and nearest-neighbor interaction $V$, in combination with the proximity to the VHS. In the simplest classical picture, or if the kinetic energy of electrons is neglected, the potential energy per site for the $\sqrt{2} \times \sqrt{2}$ charge configuration in Fig. 4(a) is: $\frac{1}{2}\left[\frac{1}{2}U(n+\delta)^2 + \frac{1}{2}U(n-\delta)^2\right] + 2V(n+\delta)(n-\delta) = \frac{1}{2}Un^2 + \left(\frac{1}{2}U - 2V\right)\delta^2$. Here $n$ is the average electron density and $\delta$ is the charge modulation. If $V > U/4$, it becomes energetically favorable for $\delta$ to be nonzero, resulting in the $\sqrt{2} \times \sqrt{2}$ charge order. This classical transition applies to a system with flat band (with infinite mass). In the realistic system, however, quantum fluctuations from itinerant electrons may melt the charge order near the classical transition point. To take quantum fluctuations into account, we perform functional renormalization group (FRG) calculations [43, 44] for an overdoped cuprate with the normal state FS shown in Fig. 4(b). The red arrow indicates scattering between the VHS saddle points, yielding logarithmically diverging susceptibilities at the momentum related to the $\sqrt{2} \times \sqrt{2}$ density wave if $U < 4V$ (or $V > U/4$). The saddle points can also lead to double-logarithmically diverging susceptibilities in the singlet-pairing channel. Such effects combined with the interactions are included and treated on equal footing in FRG method. The model has been extensively discussed in the literature but mainly for Hubbard interaction only [45, 46]. Here we are interested in the proximity to the CDW phase, hence we further include a nearest-neighbor repulsion $V$. Figure 4(c) displays the energy scale $\Lambda_c/t$ (representative of transition temperature) for the various orders as a function of $V$ for a fixed $U = 3t$, where $t$ is the nearest-neighbor hopping integral (we also include a next-nearest hopping $t' = -0.3t$). We find that the key factor determining which phase is realized is the ratio $V/U$, and thanks to the proximity to the VHS the transition point is pushed only slightly above the classical limit $V/U = 1/4$. Below the transition point, the CDW becomes dynamic and such fluctuations are seen to drive an extended $s$-wave superconducting state. As $V$ becomes even smaller, the previous studies [45, 46] show $d$-wave pairing becomes the leading order. The question is whether $V/U$ can reach and go beyond the transition point. Recent X-ray absorption spectra [47] show that with increasing doping the doped holes start entering the Cu $3d$ states, hence the effective local $U$ decreases (while the longer-ranged $V$ is less influenced). Therefore, the $\sqrt{2} \times \sqrt{2}$ CDW phase is expected to be realized in the overdoped regime, in consistency with the experimental observation.

In summary, STM experiments on a strongly overdoped non-SC Bi-2201 reveal the existence of dispersive quasiparticle interference patterns and nanoscale patches of static $\sqrt{2} \times \sqrt{2}$ charge order, which illustrate the distinct electronic structure of the heavily overdoped regime. We provide a theoretical model and FRG calculations for the charge order in terms of enhanced nearest-neighbor Coulomb repulsion $V$ versus the local Hubbard repulsion $U$, as well as the proximity to the VHS. These results shed important new lights on the unique electronic structure and electronic order in strongly overdoped cuprates.

**Acknowledgments** This work is supported by NSFC and MOST of China. XJZ thanks financial support from the National Key Research and Development Program of China (2016YFA0300300) and the Strategic Priority Research Program b of the Chinese Academy of Sciences (Grant No. XDB07020300).

**Figure Captions**

**Figure 1.** Schematic phase diagram and STM results on the overdoped non-SC Pb-Bi2201. (a) The schematic electronic phase diagram of Bi-2201, and the red arrow indicates the approximate location of the sample studied here. (b) Atomically resolved topographic image acquired over an area of 350 × 350 Å$^2$. (c) Representative $dI/dV$ spectra taken at locations indicated in (b) by the colored dots. (d) The $dI/dV$ spectrum with sharp VHS peak (black curve) can be fitted well by using the function $a + b \log |E\text{-}E_{VHS}|$ with $E_{VHS}$ = 3.5 mV (red curve). (e) Spatially averaged spectrum of 256 × 256 $dI/dV$ spectra measured in the area of (b).

**Figure 2.** The dispersive QPI patterns and theoretical simulations. (a)-(c) The $dI/dV$ maps measured in the area of Fig. 1(b) at three different bias voltages +40 mV (a), -10 mV (b), and -60 mV (c) respectively. (d)-(f) Fourier transforms of the $dI/dV$ maps in Figs. 3(a)-3(c) showing the change of FS topology. The solid white circles indicate the Bragg peaks of the (Bi,Pb) atoms. (g)-(h) The calculated QPI patterns agree well with the STM data in Figs. 3(d)-3(f). (j)-(l) CECs for the bias energies used in the theoretical simulations of QPI patterns. The CEC evolves from a ($\pi$, $\pi$)-centered hole-like pocket to a (0, 0)-centered electron-like pocket as the bias decreases.

**Figure 3.** The static charge order structure. (a) The *dI/dV* map measured in the area of Fig. 1(b) at the Fermi energy (zero bias). The yellow and green dashed squares indicate charge ordered patches. (b), (c) Atomically resolved topography and *dI/dV* map acquired over a 96 × 96 Å$^2$ area (green dashed square in (a)). Clear $\sqrt{2} \times \sqrt{2}$ charge order patches are observed in the white and red dashed squares. (d)-(g) The *dI/dV* maps of the red dashed area in (b) and (c) obtained at biases 0 mV (d), -5 mV (e), -10 mV (f), and -20 mV (g). The red dots indicate the same atom-position at different biases, demonstrating the lack of dispersion of the charge order.

**Figure 4.** Theoretical calculations for the charge order. (a) Schematic drawing of a $\sqrt{2} \times \sqrt{2}$ charge modulation. (b) The normal state FS used in the FRG calculations. The red arrow indicates the scattering between the VHS saddle points. (c) Theoretical phase diagram as a function of *V/t* for a fixed $U = 3t$, showing the evolution between the charge ordered and SC phases.

# Figure 1

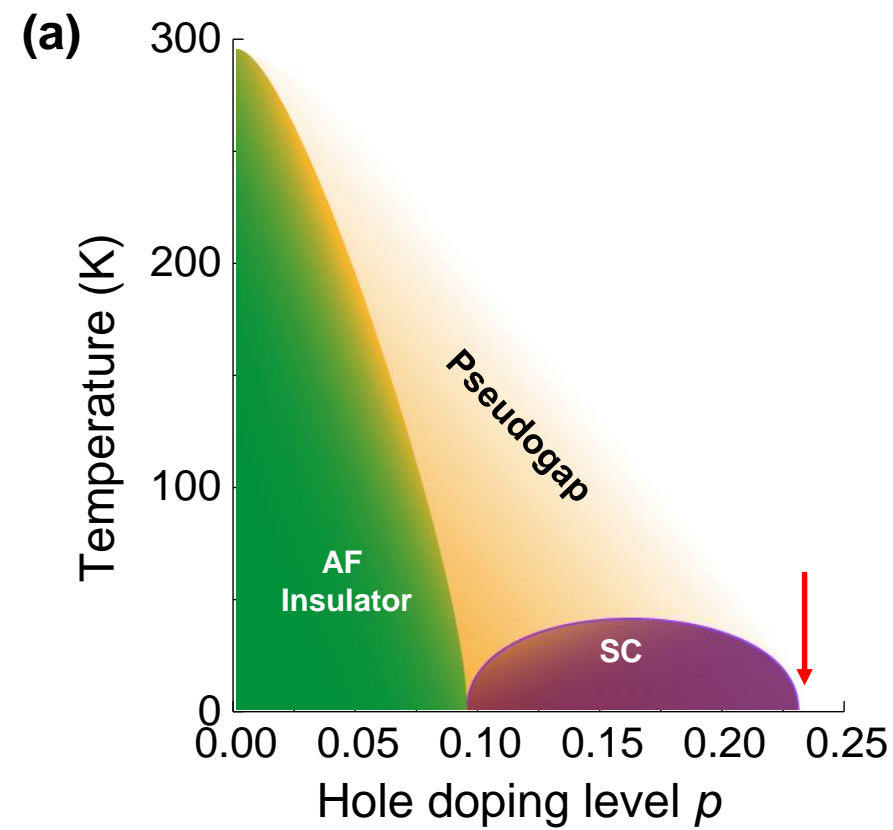
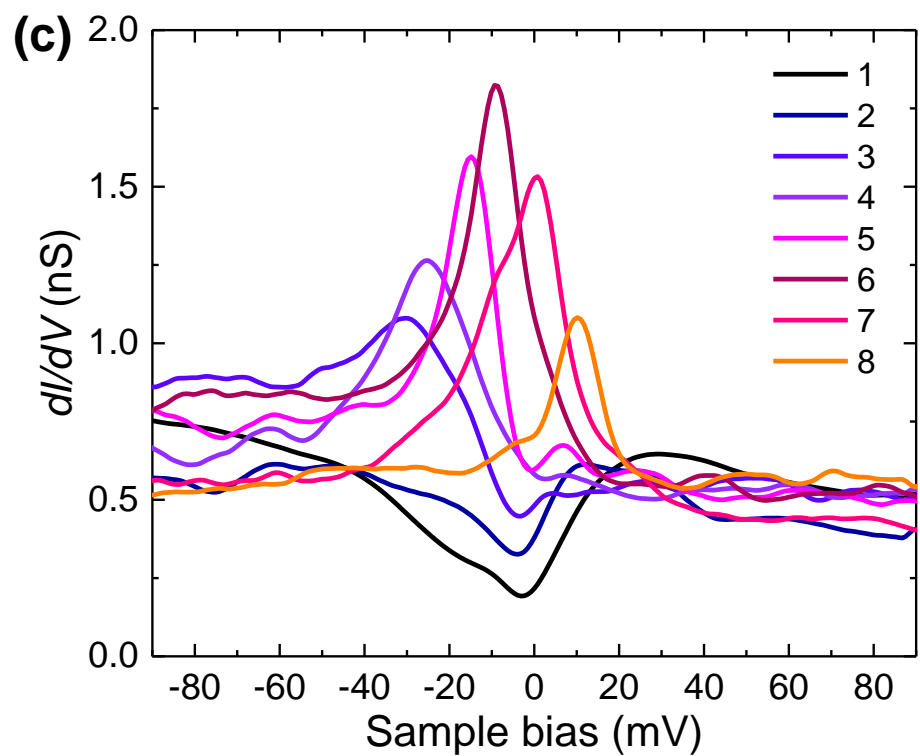
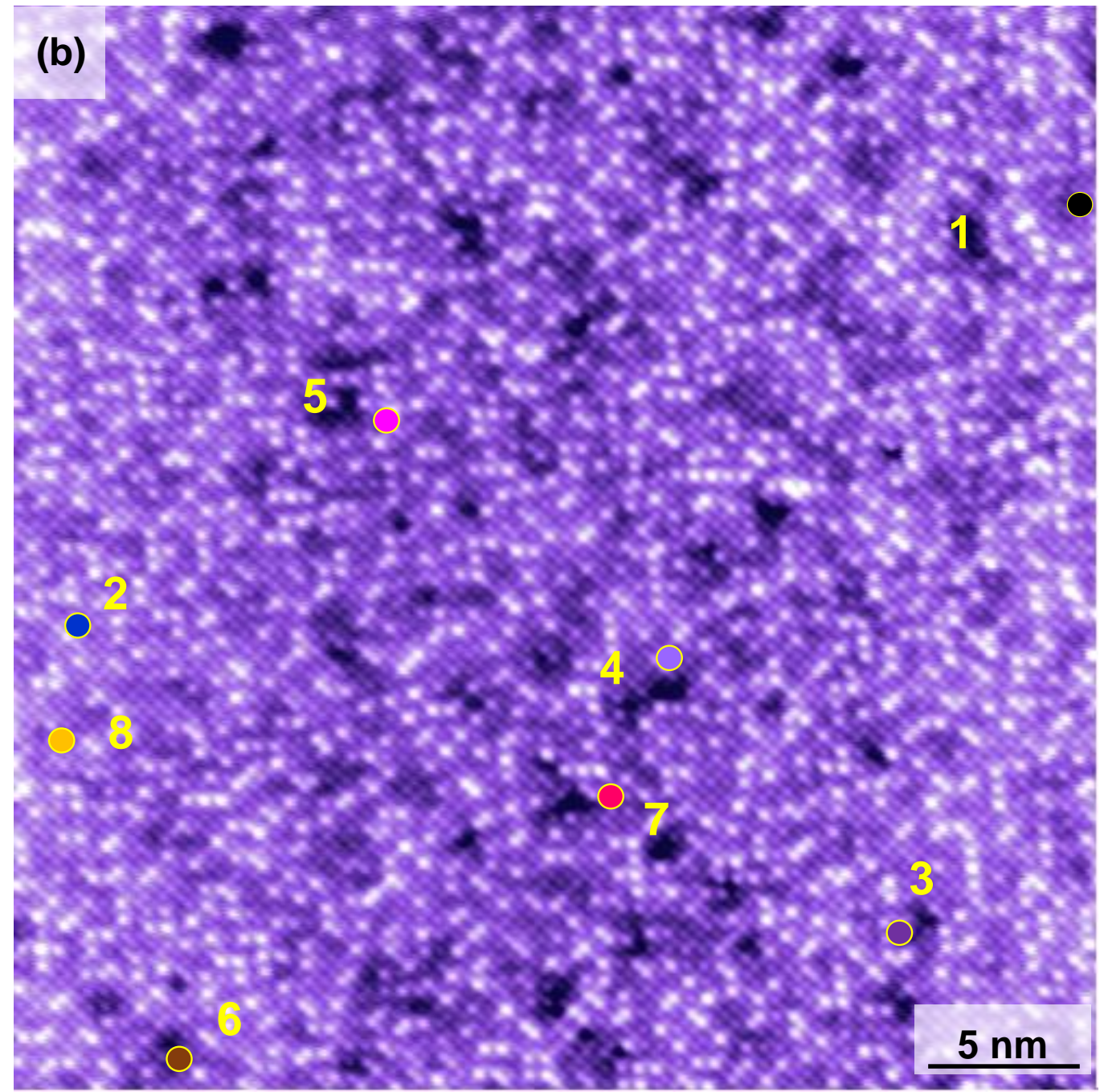
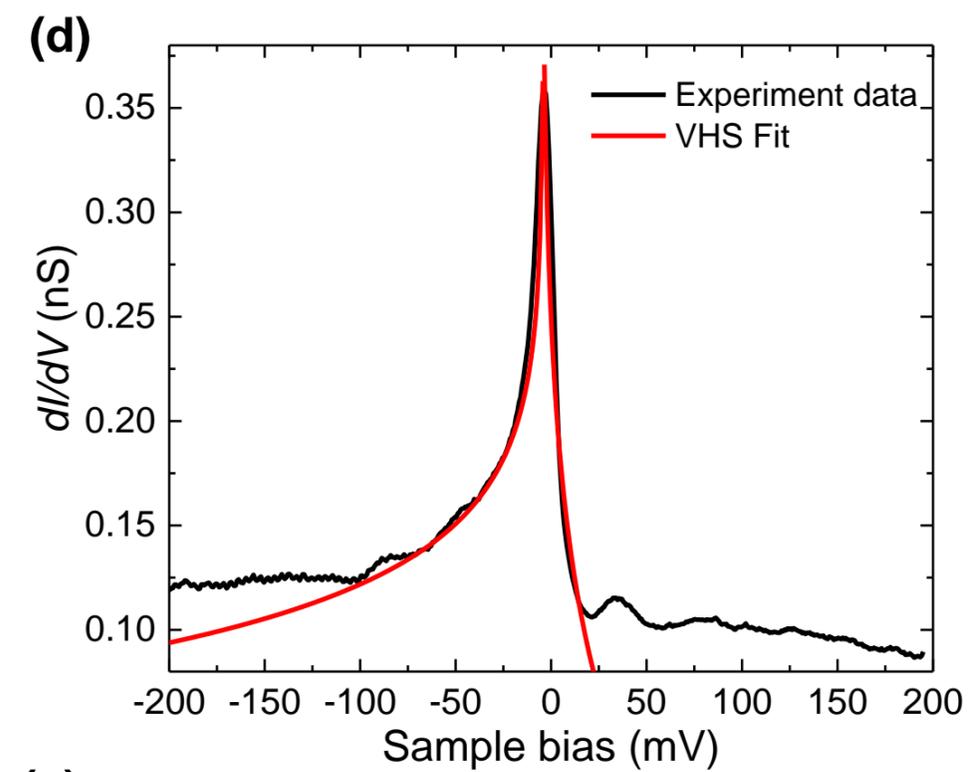
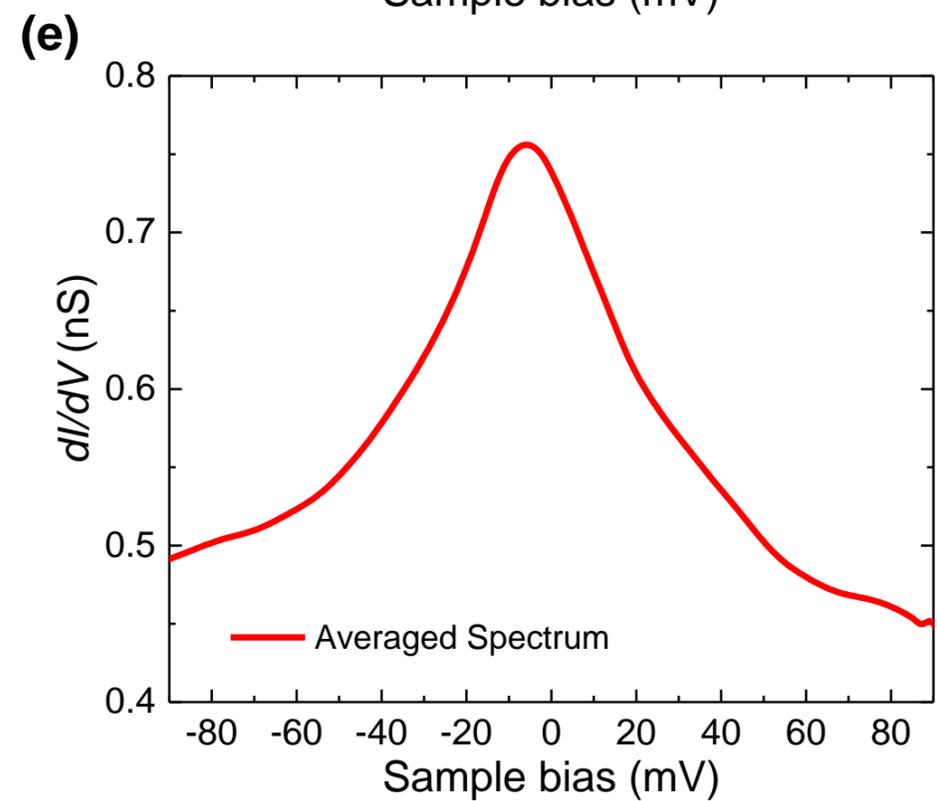

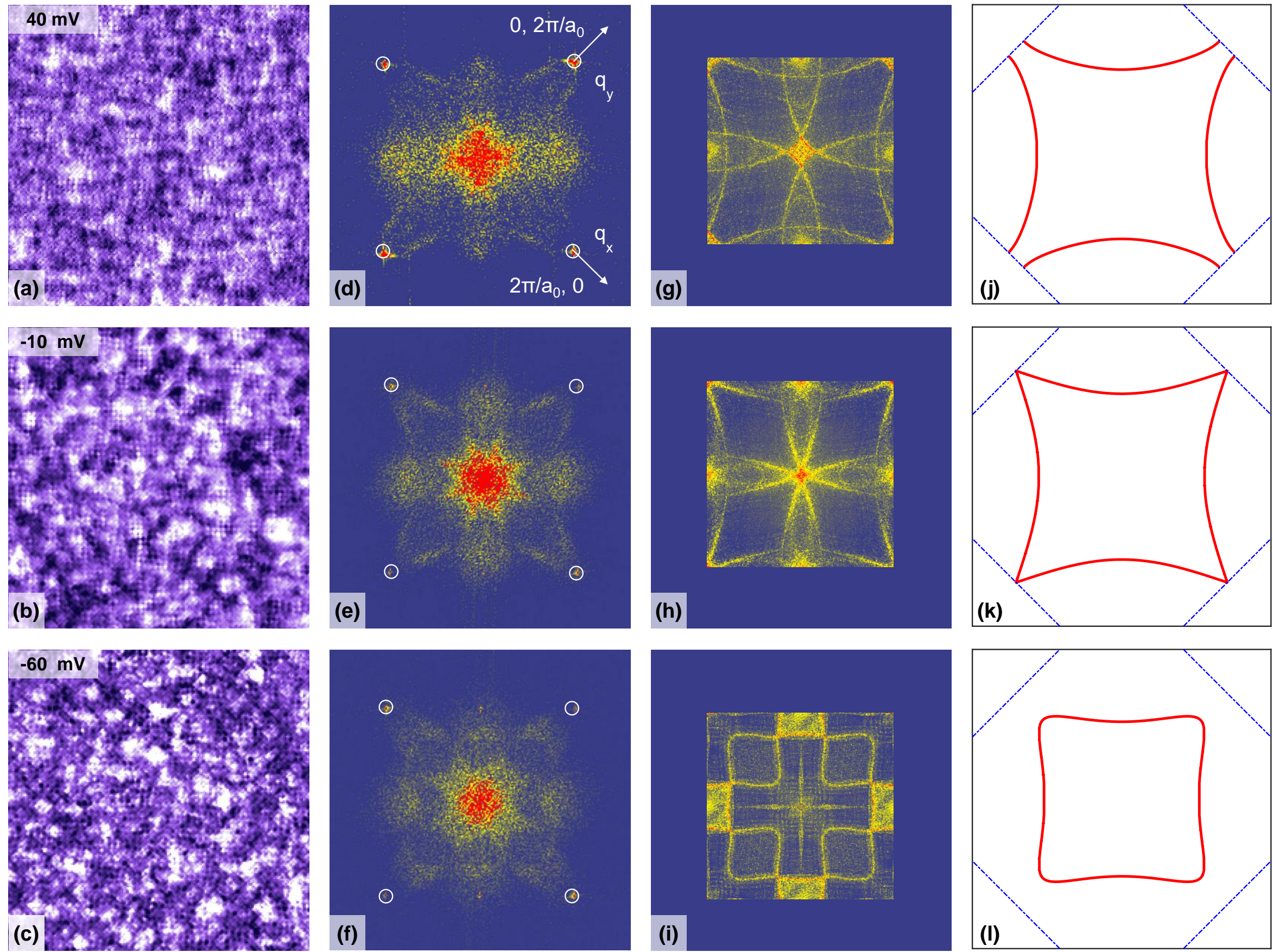

**Figure 3**

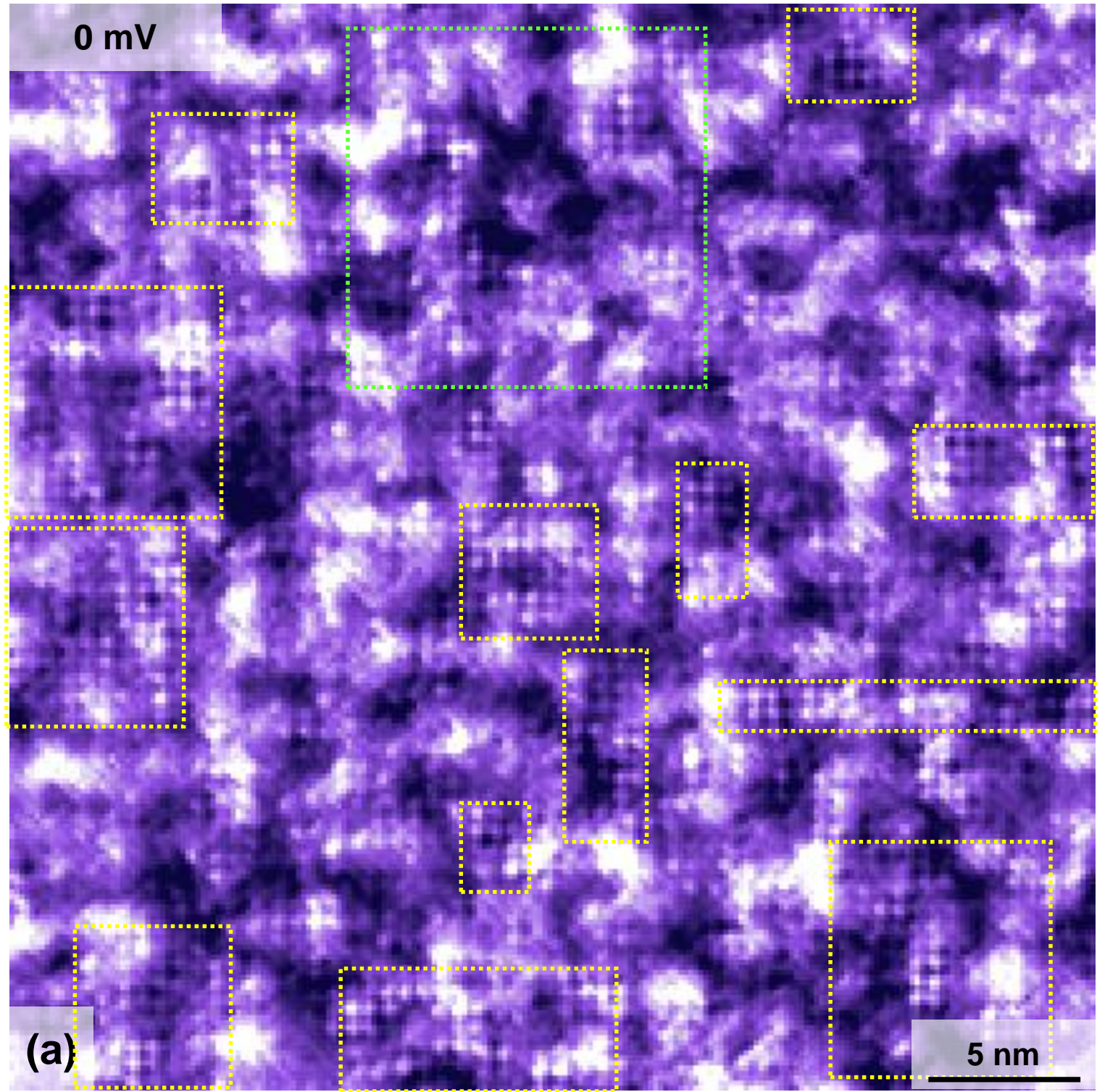
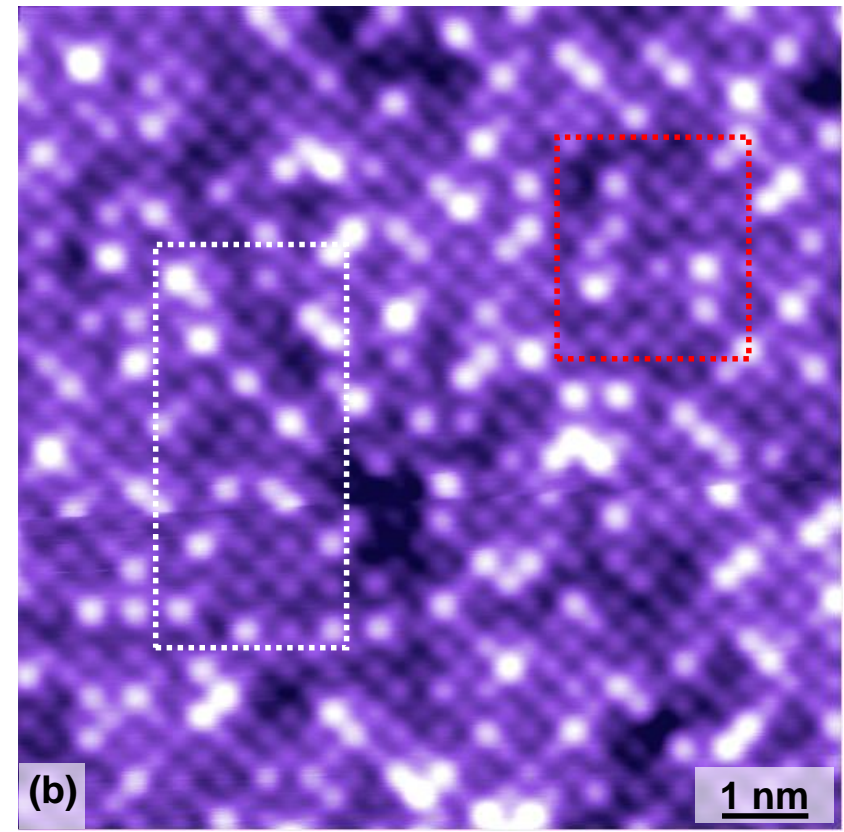
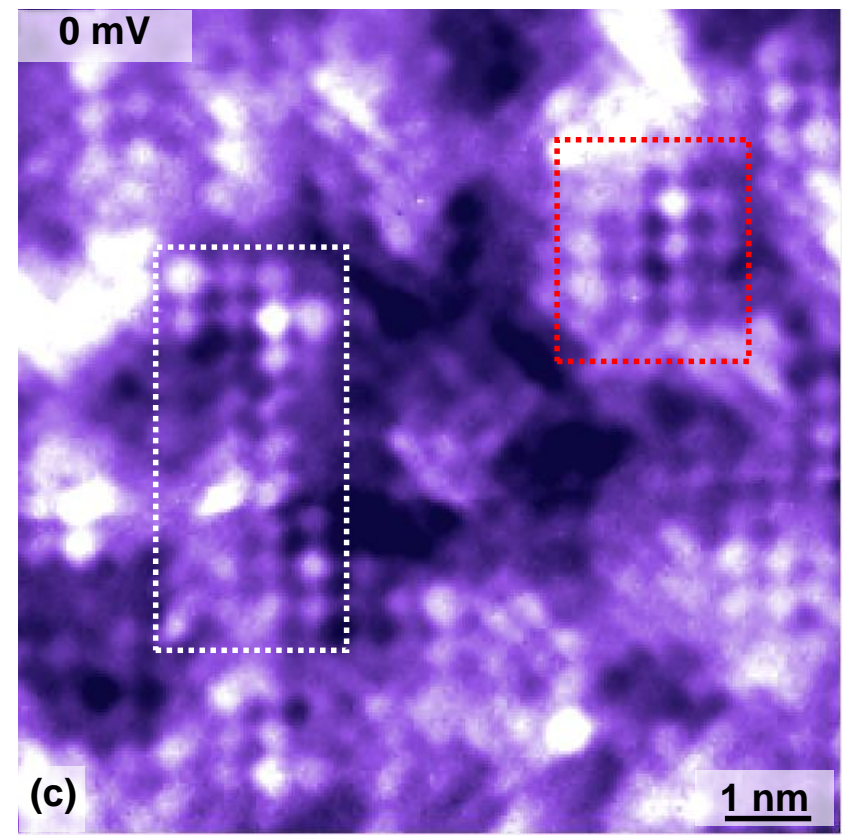
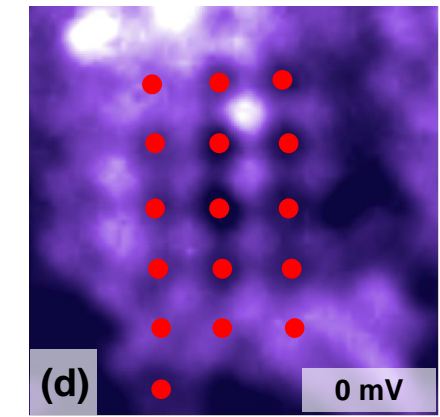
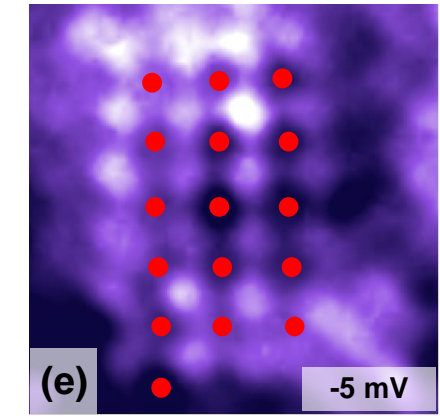
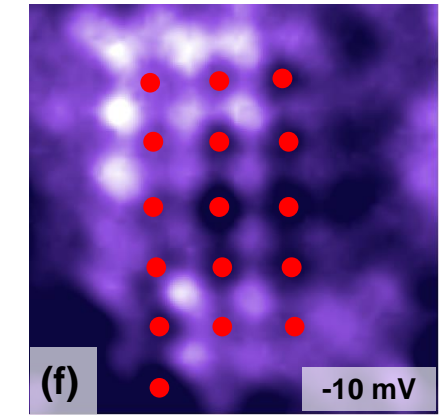
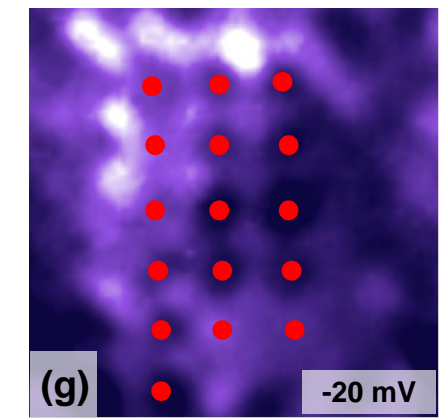

**Figure 4**

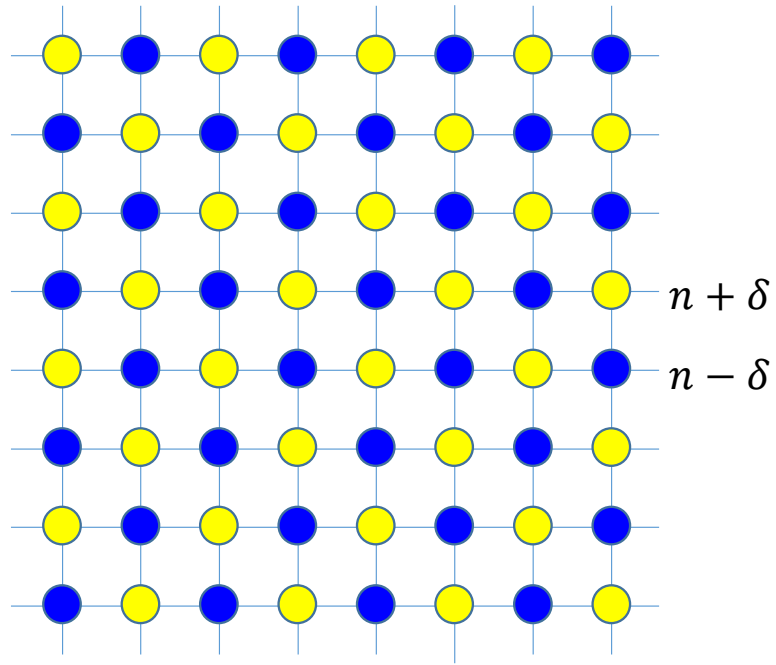

(a)

$n + \delta$
$n - \delta$

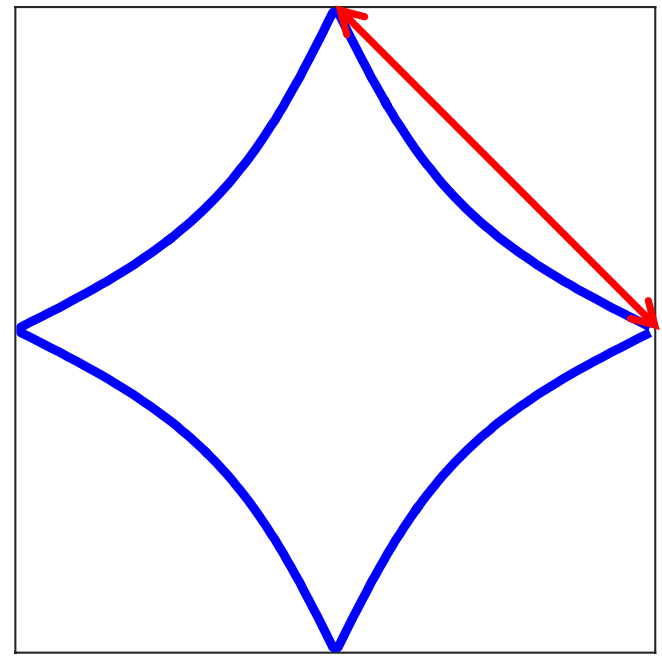

(b)

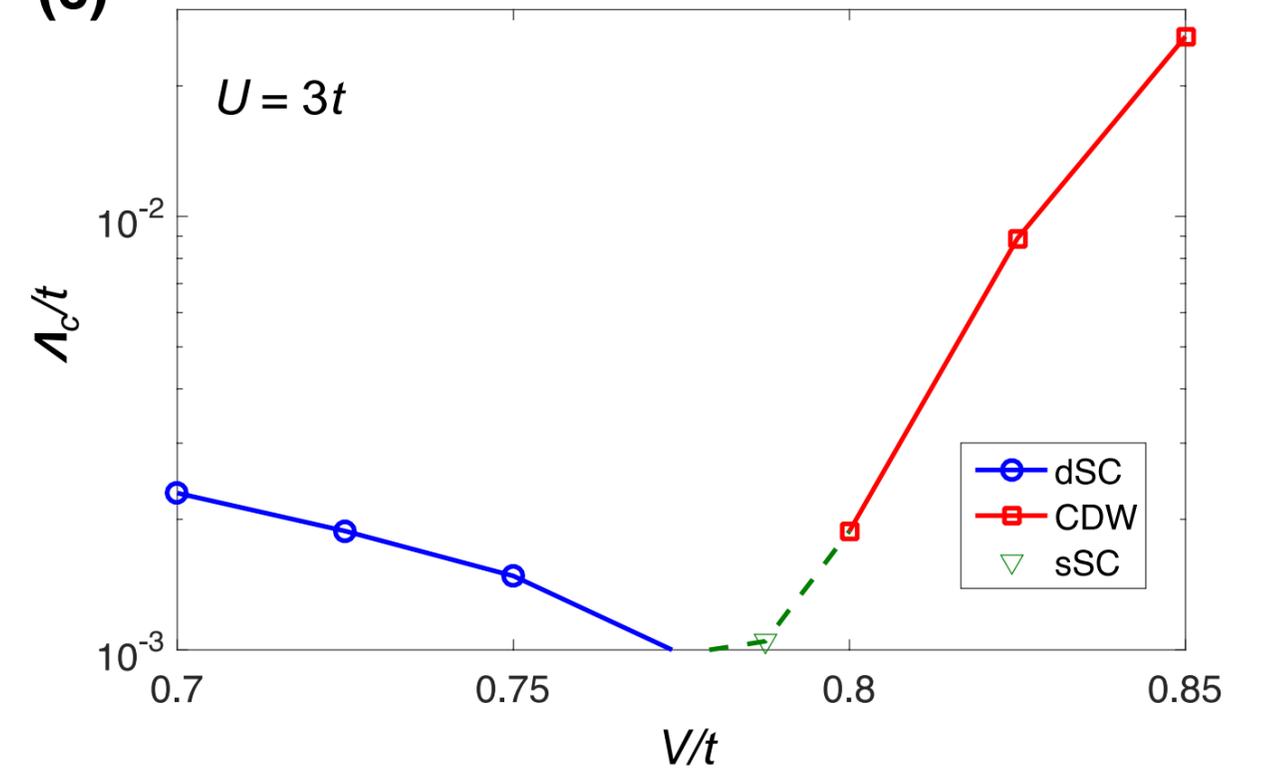

(c)

$U = 3t$

$\Lambda_c/t$

$V/t$

dSC
CDW
sSC